\documentclass[aps,pre,epsf,superscriptaddress,amsmath,amssymb,amsfonts,twocolumn,showpacs]{revtex4-1}

\usepackage{graphicx}
\usepackage{epsfig}
\usepackage{dcolumn}
\usepackage{bm}
\usepackage{braket}
\usepackage{amsmath}
\usepackage{color}

\begin{document}

\title{Bosonic quantum dynamics following a linear interaction quench\\ in finite optical lattices of unit filling}

\author{S.I. Mistakidis}
\affiliation{Zentrum f\"{u}r Optische Quantentechnologien,
Universit\"{a}t Hamburg, Luruper Chaussee 149, 22761 Hamburg,
Germany}
\author{G.M. Koutentakis}
\affiliation{Zentrum f\"{u}r Optische Quantentechnologien,
Universit\"{a}t Hamburg, Luruper Chaussee 149, 22761 Hamburg,
Germany}\affiliation{The Hamburg Centre for Ultrafast Imaging,
Universit\"{a}t Hamburg, Luruper Chaussee 149, 22761 Hamburg,
Germany}
\author{P. Schmelcher}
\affiliation{Zentrum f\"{u}r Optische Quantentechnologien,
Universit\"{a}t Hamburg, Luruper Chaussee 149, 22761 Hamburg,
Germany} \affiliation{The Hamburg Centre for Ultrafast Imaging,
Universit\"{a}t Hamburg, Luruper Chaussee 149, 22761 Hamburg,
Germany}

\date{\today}

\begin{abstract}
The nonequilibrium ultracold bosonic quantum dynamics in finite optical lattices of unit filling following a linear interaction quench from a 
superfluid to a Mott insulator state and vice versa is investigated. 
The resulting dynamical response consists of various inter and intraband tunneling modes. 
We find that the competition between the quench rate and the interparticle repulsion leads to a resonant dynamical response, at moderate ramp times, 
being related to avoided crossings in the many-body eigenspectrum with varying interaction strength. 
Crossing the regime of weak to strong interactions several transport pathways are excited.   
The higher-band excitation dynamics is shown to obey an exponential decay possessing two distinct time scales with varying ramp time. 
Studying the crossover from shallow to deep lattices we find that for a diabatic quench the excited band fraction decreases, 
while approaching the adiabatic limit it exhibits a non-linear behavior for increasing height of the potential barrier.  
The inverse ramping process from strong to weak interactions leads to a melting of the Mott insulator and possesses negligible higher-band 
excitations which follow an exponential decay for decreasing quench rate. 
Finally, independently of the direction that the phase boundary is crossed, we observe a significant enhancement of the 
excited to higher-band fraction for increasing system size. \\

\end{abstract}

\maketitle

\section{Introduction}

During the past two decades, ultracold atoms in optical lattices emerged as a versatile system to investigate many-body (MB) phenomena \cite{Lewenstein,Bloch,Yukalov,Lewenstein1}.   
A prominent example is the experimental observation of the superfluid (SF) to Mott insulator (MI) quantum phase transition \cite{Greiner,Greiner1} which, among others, 
demonstrated a pure realization of the Bose-Hubbard model \cite{Fisher,Jaksch}.  
Moreover, lattice systems constitute ideal candidates for studying nonequilibrium quantum phase 
transitions \cite{Zurek3,Polkovnikov1,Chowdhury,Sengupta,Mukherjee,Dutta3}, where  
a number of defects, induced by time-dependent quenches \cite{Polkovnikov,Andrei}, appear in the time evolving state. 
The Kibble-Zurek mechanism of such defect formation \cite{Kibble,Campo,Polkovnikov,Dziarmaga,Navon}, 
originally addressed in the context of classical phase transitions \cite{Kibble1,Zurek1}, has been tested in  
different ultracold MB settings \cite{Lamporesi,Anquez,Clark1,Chen,Meldgin,Aidelsburger} and refers to the rate 
of topological defect formation induced by quenches across phase transitions. 

Quench dynamics of ultracold bosons confined in an optical lattice covering the MI-SF transition in both directions  
has been vastly used to examine both the Kibble-Zurek mechanism \cite{Dziarmaga,Clark1,Chen,Uhlmann,Braun,Dziarmaga1,Gardas} and the approach to the adiabatic response  
limit \cite{Clark,Uhlmann,Bernier,Hung,Huber,Vidmar,Yanay,Kollath1,Gemelke,Scalettar,Sengupta1,Natu1,Cucchietti}. 
Referring to the Kibble-Zurek mechanism, recent studies \cite{Gardas,Dziarmaga1} of a linear quench across the MI-SF and back in the one-dimensional 
Bose-Hubbard model demonstrated that the excitation density and the correlation length satisfy the Kibble-Zurek 
scaling for limited ranges of quench rates. 
On the other hand and focussing on the slow quench dynamics across the MI-SF transition many recent works evinced the formation 
and melting of Mott domains \cite{Bernier,Vidmar,Gemelke,Scalettar,Sengupta1}, 
the growth of interparticle correlations \cite{Yanay,Kollath1,Braun,Cucchietti,Navez} and the consequent equilibration process \cite{Hung,Andrei,Natu1}.  

Despite the enormous theoretical and experimental efforts in this field, the response of such systems subjected to abrupt or quasi-adiabatic quenches 
has not been completely understood and deserves further investigation. 
In particular, a nonadiabatic quench inevitably excites the system, and a number of defects including higher-band excitations can 
be formed during the dynamics. 
The latter implies the necessity to consider a multiband treatment \cite{Lacki} of the nonequilibrium correlated dynamics and to obtain information about the  
higher-band excitation spectrum \cite{Kinnunen,Proukakis,Alon_lr,Alon_lr1,Theisen,Beinke} being inaccessible by the lowest Bose-Hubbard model or mean-field (MF) methods.  
Promising candidates for such investigations constitute few-body systems \cite{Blume,Blume1} being accessible by current state of 
the art experiments \cite{Zurn,Serwane,Wenz,Kaufman}. 
In this context, it is possible to track the microscopic quantum mechanisms \cite{Mistakidis,Mistakidis1,Mistakidis2,Jannis} 
consisting of intraband and interband tunneling, namely tunneling within the same or between energetically different single-particle bands respectively, 
and to avoid finite temperature effects.  
Such few-body systems do not serve as a platform to confirm the Kibble-Zurek scaling hypothesis due to their finite size \cite{Cucchietti}.  
However, they provide useful insights into the largely unexplored scaling of few-body defect density including the formation and melting of Mott domains 
and the excited to higher-band fraction participating in the dynamics.  

In the present work we consider few bosons confined in an optical lattice of unit filling.  
Thereby, the ground state for increasing interaction strength experiences the few-body analogue of the SF to MI transition.  
We first analyze the MB eigenspectrum for varying interparticle repulsion, revealing the existence of narrow and wide avoided crossings 
between states of the zeroth and first excited band. 
Then, we apply a linear interaction quench (LIQ) protocol either from weak to strong interactions (positive LIQ) or inverserly (negative LIQ) covering in both cases 
the diabatic to nearly adiabatic crossing regimes. 
As a consequence we observe a dynamical response consisting of the lowest band tunneling and higher-band excitations. 
Overall, we find an enhanced dynamical response at moderate quench rates rather than in the abrupt or almost adiabatic regimes. 
The lowest band dynamics consists of first and second order tunneling \cite{Folling,Meinert,Chen1}.    
These modes can be further manipulated by tuning either the interaction strength after the quench (postquench interaction) or the 
height of the potential barriers in the optical lattice.  
Furthermore, we show that following a positive LIQ the excited to higher-band fraction obeys a bi-exponential decay for varying ramp time. 
The latter decay law possesses two time scales being related to the width of the existing avoided crossings in the eigenspectrum. 
However, the interband tunneling \cite{Cao1,Cao2},  
with varying height of the potential barrier exhibits a more complex behavior.   
For diabatic quenches it decreases, while for smaller quench rates it scales non-linearly possessing a maximum at a certain height of the potential barrier.    
The latter behavior manifests the strong dependence of the excited to higher-band fraction on the quench rate. 
Moreover, the excited fraction for a varying postquench interaction strength features different scaling laws.   
Approaching the region of the corresponding avoided crossing it exhibits a non-linear growth, while for stronger interactions it increases 
almost linearly. 
On the contrary, for a negative LIQ we observe the melting of the MI. 
Here, the lowest band transport (intraband tunneling) is reduced when compared to the inverse scenario, while the 
excited fraction is negligible obeying an exponential decay both with varying ramp time and potential height. 
Finally, for both positive and negative LIQs the higher-band fraction is significantly enhanced for increasing system size. 

This work is organized as follows.  
In Sec. \ref{theory} we introduce our setup and outline the multiband expansion being used for a microscopic analysis of the dynamics. 
Sec. \ref{dynamics} presents the resulting dynamics induced by a LIQ connecting the weakly to strongly correlated regimes and back in a triple 
well of unit filling.   
To extend our findings in Sec. \ref{fivewells} we discuss the LIQ dynamics for larger lattice systems of unit filling.  
We summarize and discuss future perspectives in Sec. \ref{conclusions}.  
Appendix A describes our computational methodology.

\section{Setup and analysis tools}\label{theory}

The Hamiltonian of $N$ identical bosons each of 
mass $M$ confined in an one-dimensional $m$-well optical lattice employing a LIQ protocol reads   
\begin{equation}
H=\sum_{i=1}^{N}\left(\frac{p_i^2}{2M}+V_0\sin^2(kx_i)\right)+g(t,\tau)\sum_{i<j}\delta(x_i-x_j). \label{Hamilt}
\end{equation} 
The lattice potential is characterized by its depth ${V_0}$ and 
periodicity $l$, with $k = 2\pi/l$ being the wave vector of the counterpropagating lasers forming the optical
lattice.   
To restrict the infinitely extended trapping potential to a finite one with $m$ wells and
length $L$, we impose hard wall boundary conditions at the
appropriate positions, $x_{m} = \pm
\frac{{m\pi }}{{2k}}$.  

Within the ultracold regime, the short-range interaction potential between particles located at positions ${x_i}$,  
can be adequately described by $s$-wave scattering.    
To trigger the dynamics we follow a LIQ protocol.        
At $t=0$ the interatomic interaction is quenched from the initial value $g_{i}$ to a final one $g_f$ in a linear manner 
for time $t \in [0,\tau ]$ and then it remains a constant $g_f$. 
Therefore, our protocol reads
\begin{equation} 
\begin{split}
\label{g_t}
g(t,\tau)=&g_{i}+\delta g \frac{t}{\tau}.
\end{split}
\end{equation} 
Here, $\delta g=g_f-g_{i}$ denotes the quench amplitude of the linear quench, while $g_i$ ($g_f$) is the effective 
one-dimensional interaction strength before (after) the quench. 
The effective one-dimensional interaction strength \cite{Olshanii} is given by 
${g_{1D}} =
\frac{{2{\hbar ^2}{a_0}}}{{Ma_ \bot ^2}}{\left( {1 - {\left|
{\zeta (1/2)} \right|{a_0}}/{\sqrt 2 {a_ \bot }}} \right)^{ -
1}}$. 
Here ${a_ \bot } = \sqrt{\hbar /{M{\omega _ \bot }}}$ is the transverse harmonic oscillator length with ${{\omega _ \bot }}$ the
frequency of the two-dimensional confinement and ${a_0}$ denotes the free space 3D $s$-wave
scattering length.  
Experimentally, the effective interaction strength can be tuned either via $a_0$ with the aid of  
Feshbach resonances \cite{Kohler,Chin} or via the corresponding transversal 
confinement frequency $\omega_\bot$ \cite{Olshanii,Kim,Giannakeas}. 

In the following, the Hamiltonian (1) is rescaled in units of the recoil
energy ${E_{\rm{R}}} = \frac{{{\hbar ^2k^2}}}{{2M}}$.  
Then, the corresponding length, time and interaction strength scales are given in units of
${k^{ - 1}}$, $\omega_{\rm{R}}^{-1}=\hbar E_{\rm{R}}^{ - 1}$ and $E_{\rm{R}}k^{-1}$, respectively. 

To simulate the nonequilibrium dynamics we employ the Multi-Configuration Time-Dependent Hartree method for
Bosons (MCTDHB) \cite{Alon,Alon1} which exploits an expansion in terms of time-dependent 
variationally optimized single-particle functions (see Appendix A for more details). 
In contrast to the MF approximation, within this approach we account for the system's interparticle correlations and  
hence we will refer to MCTDHB simply as the MB approach.  
However, for the analysis of the induced dynamics in lattice systems, it is more intuitive to rely on a time-independent 
MB basis instead of a time-dependent one.  
Here, we project the numerically obtained wavefunction on a time-independent 
number state basis being constructed by the single-particle Wannier states localized on each lattice site. 
The MB bosonic wavefunction of a system with $N$ bosons, $m$-wells and $j$ localized single
particle states \cite{Mistakidis,Mistakidis1} reads 
\begin{equation}
\label{expansion}
\left| \Psi  \right\rangle  = \sum\limits_{\{\vec{n}_{i}\}}
{{C_{\{\vec{n}_i\}}}{{\left| {{\vec{n}_1},{\vec{n}_2},...,{\vec{n}_m}} \right\rangle
}}},
\end{equation}
where ${{\left| {{\vec{n}_1},{\vec{n}_2},...,{\vec{n}_m}} \right\rangle}}$ is the multiband Wannier number state, the element 
${{\vec{n}_i}}=\ket{n_i^{(1)}}\otimes\ket{n_i^{(2)}}\otimes....\otimes\ket{n_i^{(j)}}$ and the Wannier orbital $\ket{n_i^{(k)}}$   
refers to the number of bosons which reside at
the $i$-th well and $k$-th band. 
For instance, in a setup with $N=3$ bosons confined in a triple well
i.e. $m=3$, which includes $k=3$ single-particle states, the state 
$\ket{1^{(0)},1^{(1)},1^{(0)}}$
indicates that in the left and right wells  
one boson occupies the Wannier orbital of the zeroth excited band while in the middle well there is one boson in the Wannier orbital   
of the first excited band. 
Below, when we refer to a boson that resides within the zeroth (ground) band we shall omit the zero index. 
Here, one can realize three different energetic classes of number states with respect to the interparticle repulsion. 
Namely, the triples $\{ {\left| {3,0,0}
\right\rangle}+\circlearrowright\}$ ($T$), 
the single pairs $\{ {\left| {2,1,0}\right\rangle}+\circlearrowright\}$ ($SP$) 
and the singles $\{ {\left| {1,1,1}\right\rangle}+\circlearrowright\}$ ($S$), 
where $\circlearrowright$ stands for all corresponding permutations. 
For later convenience, on the analysis part, we further classify the excited band energetic classes into  
single-particle excitation ($SE$) and higher excited ($HE$) classes. 
The former [latter] class involves states of single [double] occupancy in every site with one 
excitation to the first excited band e.g. $\{\ket{1,1^{(1)},1}+\circlearrowright\}$ [$\{ \ket{1 \otimes 1^{(1)},1,0}+\circlearrowright\}$ 
and $\{ \ket{1^{(1)},2,0}+\circlearrowright\}$].  
 
Moreover, according to the above expansion the time averaged probability for bosons to lie in a higher single-particle band reads
\begin{equation}
\label{excitation}
\bar{P}_{exc}(\tau)=1-\frac{1}{T}\int_0^T dt~ P_0(t;\tau).   
\end{equation}
Here, $P_0(t;\tau)=\sum_{\{n_i\}}|\braket{n_1,n_2,n_3|\Psi(t)}|^2,~ i=1,2,3$ denotes the probability for  
all particles to reside within the zeroth band, while $T$ refers to the considered finite evolution time where $\bar{P}_{exc}(\tau)$ 
has converged to a certain value. 
The above probability amplitude will be a main tool for the investigation of the scaling of the excited to higher-band fraction with respect to $\tau$.

\section{LIQ dynamics in a triple well of unit filling}\label{dynamics}
 
To analyze the LIQ induced dynamics of our system, it is instructive first to demonstrate the dependence of the eigenstates on the interparticle repulsion.   
Thus, we first investigate the eigenspectra of the system with varying interaction strength [Sec. \ref{eigen}],  
which are subsequently related to the LIQ dynamics [Secs. \ref{SFtoMOTT} and \ref{MOTTtoSF}].  

\subsection{Eigenspectra} \label{eigen}

In contrast to the discrete Bose-Hubbard model, here we employ a continuum space Hamiltonian [see Eq. (\ref{Hamilt})], which allows us to
resolve quench induced higher-band excitations \cite{Koutentakis}.  
For completeness we note that the Bose-Hubbard model is adequate for the theoretical description of the quench dynamics  
in deep lattices (i.e. large $V_0$) and for relatively small quench amplitudes when compared to the band gap.  
To expose the underlying physical processes that lead to the emergence of such MB excited states \cite{Alon_lr,Alon_lr1,Theisen,Beinke} 
we examine, below, the eigenenergies of three particles confined in a triple well potential as a function of the interaction strength $g$.  
\begin{figure}[h]
    \centering
    \includegraphics[width=1.0\columnwidth]{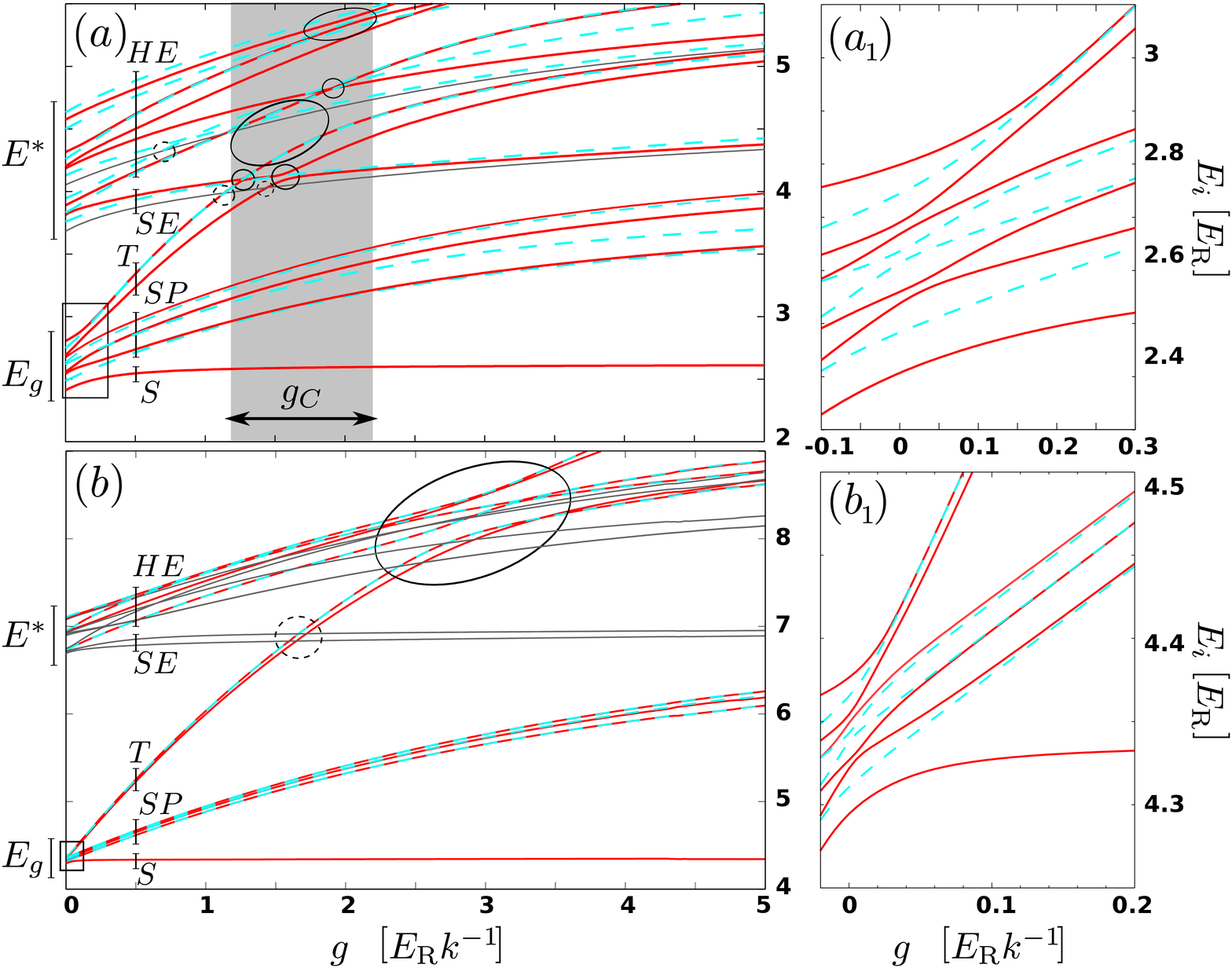}
    \caption{Dependence of the lowest 25 eigenenergies $E_i$ on the interaction strength $g$. 
    The system consists of three bosons confined in a triple well with potential depth ($a$) $V_0=4$ and ($b$) $V_0=10$.   
    Solid (dashed) lines represent parity even (odd) $E_i$'s.   
    Wide (narrow) avoided crossings possessing a width $\delta E > 0.01$ ($\delta E < 0.01$) are denoted by solid (dashed) circles.  
    The eigenenergies of the eigenstates that do not contribute to any wide avoided crossing are shown in grey.  
    The energy regions $E_g$, $E^*$ and the subbands $S$, $SP$, $T$, $SE$ and $HE$ are marked by the respective bars. 
    The grey scaled interaction interval $g_C$ in ($a$) denotes the region of avoided crossings between the $SE$ and $T$ as well 
    as the $T$ and $HE$ classes for the case of a shallow lattice $V_0=4$. 
    The solid boxes indicate the SF to MI transition.
    For better visibility of the avoided crossings in the vicinity of $g=0$ (see solid boxes), ($a_1$) and ($b_1$) provide 
    the lowest 10 eigenenergies of ($a$) and ($b$) respectively.}
    \label{fig:eigenspectra}
\end{figure}

First we focus on the case of a relatively shallow triple well, namely $V_0=4.0$, see Fig. \ref{fig:eigenspectra} ($a$).   
For very small interactions, $g \simeq 0$, the MB eigenstates are energetically
categorized according to their corresponding particle configuration in terms of single-particle bands.  
For instance, Fig. \ref{fig:eigenspectra} ($a$) shows that the eigenstates of the system are predominantly bunched onto two energy regions denoted by $E_g$ and $E^*$ respectively.     
The eigenstates lying within $E_g$ possess zero higher-band excitations, while those bunched onto $E^*$ refer to states with one single-particle 
excitation to the first excited band. 
The width of the aforementioned energy regions (bandwidth) depends on the tunneling coupling between the different sites. 
Note here that the term tunneling coupling refers to the corresponding inverse tunneling rate \cite{Fisher,Jaksch}.   
The distance between $E^*$ and $E_g$ (band gap) is characterized by the band gap between the ground and the first excited band of the non-interacting system. 
Regarding the decomposition of each eigenstate in terms of (localized) Wannier number states [see Eq. (\ref{expansion})] it turns out 
that it is an admixture of all the energetic classes $S$, $SP$ and $T$.   
The latter indicates the spatial delocalization of the bosons within the triple well, and therefore manifests the few-body analogue of the SF phase for low interactions.

For increasing interaction strength the energy expectation value of the number states belonging to the $SP$ and $T$ classes increases strongly.   
The same holds for the eigenenergies of the eigenstates to which the aforementioned number states are contributing. 
For $0< g \lessapprox 0.5$ several avoided crossings are observed, see Fig. \ref{fig:eigenspectra} ($a_1$). 
These avoided crossings manisfest the tunneling coupling between the $S$, $SP$ and $T$ number states of the same parity \cite{Koutentakis,Cao1}.  
The aforementioned interaction regime corresponds in our few-body system to the region where the transition from the SF to the MI phase takes place.  
For $g \ge 0.5$ the eigenenergies of the lowest band become well separated into three subbands
according to the energetic class of their dominant number state, see Fig. \ref{fig:eigenspectra} ($a$). 
The ground state of the system is dominated by the $S$ class manifesting the few-body analogue of the MI phase. 
The $SP$ and $T$ class dominated eigenstates are also bunched together forming the $SP$ and $T$ subbands.  
Moreover, the eigenstates of the $T$ subband (being the most sensitive to interparticle repulsion) experience 
wide (see solid circles) and narrow (see dashed circles) \cite{Note1} avoided crossings  
with the eigenstates possessing a higher-band excitation.   
The wide avoided crossings are related to the onset of the cradle mode \cite{Mistakidis,Mistakidis1} 
and are a consequence of the interaction induced decay of an $SP$ or $T$ state caused by the scattering of one of 
the bosons that reside in the same well to the first excited state of an adjacent site \cite{Note2}. 
This latter process gives rise to the so-called cradle mode which represents a dipole-like intrawell oscillation 
in the outer wells of the finite lattice (for a detailed description on the generation and properties of this mode see \cite{Mistakidis,Mistakidis1}).  

The same overall behavior can be observed for deeper lattices, see for instance in Fig. \ref{fig:eigenspectra} ($b$) the case of $V_0=10$.  
As shown, the differences between shallow and deep lattices are mainly quantitative \cite{Note3}.    
The bandwidth of each subband decreases as a consequence of the reduced tunneling coupling, while the corresponding subband energy gap increases.  
Most importantly, the transition from the SF to the MI state is realized for smaller interactions when compared to the case of $V_0=4$ 
[see in particular Figs. \ref{fig:eigenspectra} ($a_1$) and ($b_1$)].  
Moreover, as a consequence of the increased band gap, the positions of the avoided crossings related to the cradle mode (see solid circles in Fig. \ref{fig:eigenspectra}) 
occur at larger interparticle repulsions $g$. 
Concluding, all differences in the eigenspectrum caused by $V_0$ can be traced back to the increased intrawell localization of the particles for deeper lattices.

In the following sections, we study the dynamical response induced by a LIQ from $g_i=0$ to $g_f=2$ and back. 
The choice of $g_i=0$ ensures that the initial state within a positive LIQ consists of an admixture of all energetic classes, while due to the 
postquench interaction, $g_f=2$, the system remains adequately detuned from the cradle mode even in the fully diabatic (i.e. abrupt) limit. 
Moreover, following a negative LIQ from $g_i=2$ to $g_f=0$ the system initially resides in a MI state and finally in an admixture of 
several lowest band states. 
Taking advantage of the above presented eigenspectra, we expect that the higher-band dynamics strongly depends on the 
ramping rate of the SF to MI transition resulting in a different population of the $T$ class (eigen)states.

\subsection{Quench dynamics from SF to MI phase} \label{SFtoMOTT}

\begin{figure}[ht]
        \centering
           \includegraphics[width=0.45\textwidth]{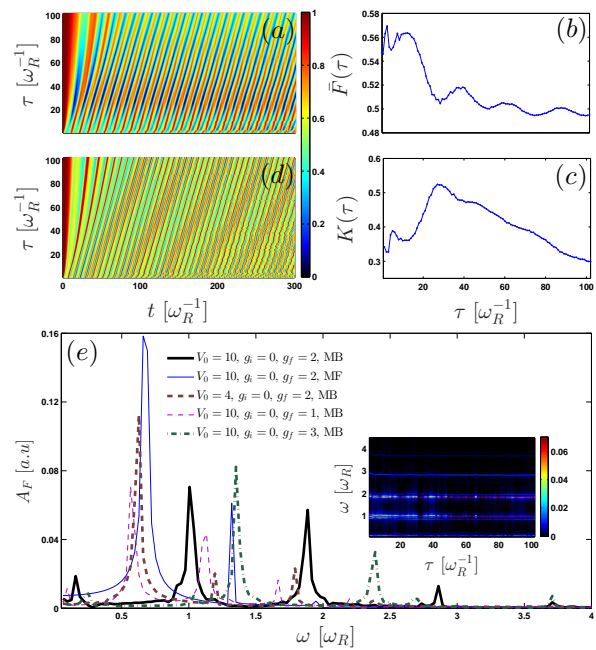}
                \caption{Fidelity evolution for varying ramp time $\tau$ after a LIQ from $g_i=0$ to $g_f=2$. 
                Shown are the cases of ($a$) a shallow $V_0=4$ and ($d$) a deep $V_0=10$ triple well.
                ($b$) Mean fidelity $\bar{F}(\tau)$ and ($c$) normalized variance of the fidelity $K(\tau)$ in the case of a shallow lattice for varying $\tau$.  
                ($e$) Spectrum of the fidelity $F(\omega)$, following a LIQ with $\tau=8$ for different barrier heights, initial and final interactions 
                within the MB approach or the MF approximation (see legend). 
                Inset: $F(\omega)$ following a LIQ from $g_i=0$ to $g_f=2$ for varying ramp time $\tau$.   
                The system consists of three initially non or weakly interacting bosons (see legend) confined in a triple well. } 
                \label{fig:fidelity_SF}
\end{figure}

Let us first focus on the positive LIQ dynamics to strong interactions, namely from $g_i=0$ to $g_f=2$, of a system consisting 
of three initially non-interacting bosons confined in a triple well. 
The initial MB state is an admixture of the available lowest band number states [see also Fig. \ref{fig:eigenspectra} ($a$)], where the main contribution 
stems from the $SP$ category due to the hard wall boundary conditions.     
To gain an overview of the system's dynamical response induced by the LIQ we employ the fidelity 
$F(t;\tau)={\left| {\left\langle {\Psi_{\tau} (0)} \right|\left. {\Psi_{\tau}(t)} \right\rangle } \right|^2}$, 
being the overlap between the time evolved and the initial (ground) state \cite{Venuti,Gorin,Campbell}.    
Fig. \ref{fig:fidelity_SF} ($a$) presents $F(t;\tau)$ for a shallow lattice ($V_0=4$) with varying ramp time $\tau$.     
As it can be seen, $F(t;\tau)$ performs oscillations in time with multiple frequencies, while for increasing ramp time the system significantly departs  
from its initial state [i.e. $F(t;\tau)$ overall decreases].  
Interestingly enough, for intermediate $\tau$'s the fidelity fluctuates more prominently, see e.g. $15<\tau<35$, indicating an enhanced dynamical response. 
The same overall response is observed for deeper lattices, see for instance Fig. \ref{fig:fidelity_SF} ($d$), but the corresponding signatures of enhanced response for  
larger $\tau$'s become more faint due to the increased energy gap in the respective MB eigenspectra [see also Fig. \ref{fig:eigenspectra}].  
To further elaborate on the overall response of the system induced by the LIQ we show in Fig. \ref{fig:fidelity_SF} ($b$) the time averaged fidelity of 
Fig. \ref{fig:fidelity_SF} ($a$) for different ramp times, namely $\bar{F}(\tau)=\frac{1}{T-\tau} \int_{\tau}^T dt F(t)$, where $T=500$.    
We observe that for increasing $\tau$, $\bar{F}(\tau)$ mainly decreases possessing also some small amplitude deformations displayed as dips in $\bar{F}(\tau)$.  
This overall decrease of $\bar{F}(\tau)$ implies that the system departs more prominently from its initial state for increasing $\tau$      
and it is a manifestation of the Landau-Zener mechanism \cite{Landau,Zener,Dziarmaga}. 
Namely, the more adiabatically we drive the system it departs stronger from its initial state.     
Finally, it is observed that for large enough $\tau$, namely $\tau>70$, $\bar{F}$ tends to a constant value which indicates a tendency to approach the adiabatic ramping rate.    
The small amplitude deformations in $\bar{F}(\tau)$ refer to different resonant response regions  
at specific intervals of $\tau$.  
These strong response regions indicate that the specific combination of $\tau$'s and quench amplitude are more efficient to cross the existing avoided crossings 
[see Fig. \ref{fig:eigenspectra}] and as a consequence to depart from the initial state. 
This latter behavior is already imprinted in $F(t;\tau)$ at the final instant of the ramping, i.e. $t=\tau$. 
Indeed, within the $\tau$ intervals that $\bar{F}(\tau)$ exhibits dips (humps) the system departs significantly (negligibly)  
from its initial state at time $t=\tau$, see also Fig. \ref{fig:fidelity_SF} ($a$).  
A careful inspection of the eigenspectra, shown in Fig. \ref{fig:eigenspectra} ($a$), indicates that for smaller $V_0$'s the gap between the different energetic subbands  
reduces and therefore the corresponding tunneling processes can be more pronounced.  
The latter suggests that for shallower [deeper] lattices the system is perturbed more efficiently [inefficiently] resulting in an enhanced [reduced] dynamical response.  
To examine in more detail the dynamical response, induced by the LIQ, we employ the normalized variance of the fidelity 
\begin{equation}
\label{defects}
 K(\tau)=\frac{\sqrt{\sigma_F^2(\tau)}}{\bar{F}(\tau)}, 
\end{equation}
where $\sigma_F^2(t)=\frac{1}{T-\tau}\int_{\tau}^T dt [F(t;\tau)-\bar{F}(\tau)]^2$ denotes the temporally integrated variance of the fidelity.  
$K(\tau)$ serves as a measure for the mean fidelity variation from its mean value and it is bounded to take values within the interval $[0,1]$.   
Then, when $K(\tau)\to 1$ [$K(\tau)\to 0$] the system possesses the maximum [minimum] fluctuation from its mean final state.  
Clearly, within the ramping intervals $\tau\in (5,10), (15,35)$ the system can be driven away from its initial state in the most prominent manner [$K(\tau)$ increases] 
and therefore the dynamical response is maximized (see also below).    

 \begin{figure}[ht]
          \centering
            \includegraphics[width=0.45\textwidth]{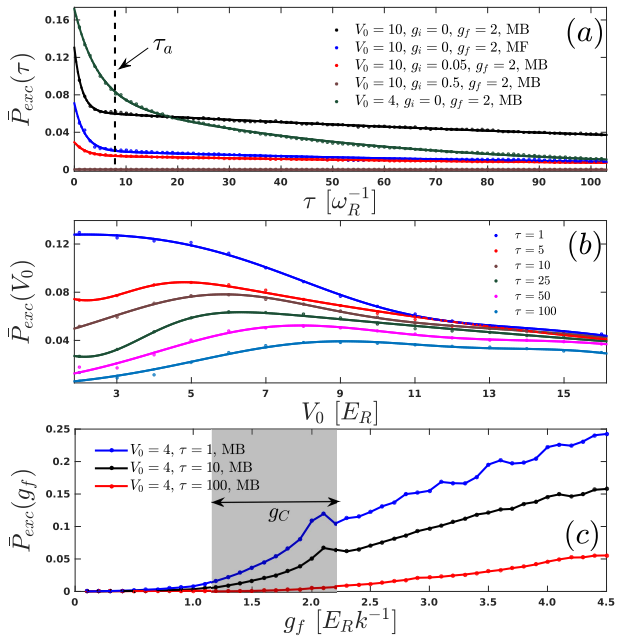}
                 \caption{Mean excitation probability of at least one boson to reside in an excited band for varying ($a$) ramp time $\tau$ of the LIQ,   
                 ($b$) barrier height $V_0$ and ($c$) postquench interaction strength $g_f$.  
                 In all cases we consider a LIQ from an initially non to a final strongly correlated state unless stated otherwise (see legends).  
                 Different curves correspond to different parameter values (see legend), while the lines belong to a numerical fitting and 
                 provide a guide to the eye. The time-scale $\tau_a$ denotes the border between diabatic and moderate ramping for $V_0=10$. 
                 The grey scaled interaction interval in ($c$) denoted by $g_C$ refers to the region of avoided crossings when $V_0=4$ between the $SE$ and $T$ as well as 
                 the $T$ and $HE$ categories, see also Fig. \ref{fig:eigenspectra} ($a$).}
                 \label{fig:excitation_SF}
 \end{figure}

To identify the participating tunneling modes induced by the positive LIQ, we inspect the fidelity spectrum \cite{Jannis,Mistakidis2}, see Fig. \ref{fig:fidelity_SF} ($e$).  
As shown the LIQ triggers five distinct tunneling modes onto the system.  
The first four modes located at $\omega_{1,2,3,4}\approx 0.2, 1, 1.8, 2.9$ for $V_0=10$, $g_f=2$ refer to intraband tunneling within the $SP$ class,  
$SP\to S$, $SP\to T$, and $T\to S$ categories respectively. 
The mode at $\omega_{5}\approx 3.75$ indicates interband interaction assisted tunneling between the $S$ and $HE$ states. 
Note here that the $SP\to S$, and $T\to S$ lowest band modes refer to second order tunneling \cite{Folling,Meinert,Chen1}, 
while the others denote single-particle transport.  
The positions of these frequency branches remain insensitive to varying $\tau$, see the inset in Fig. \ref{fig:fidelity_SF} ($e$). 
It is also worth mentioning at this point that the dominant dynamical modes $SP\to S$ and $SP\to T$ are responsible for the maximized 
dynamical response [see Fig. \ref{fig:fidelity_SF} ($c$)] within the ramping intervals $\tau\in (5,10)$ and $(15,35)$ respectively. 
To highlight the correlated MB character of the above mentioned tunneling processes we also show the corresponding fidelity spectrum 
calculated within the MF approximation. 
Here, we observe that the system features only the first two tunneling modes, namely transport within the $SP$ and $SP\to S$ categories.  
These mode frequencies are also positively shifted namely $\omega'_{1,2}\approx 0.65, 1.35$ when compared to the MB approach.    
As a next step, let us demonstrate possible control protocols of the tunneling dynamics.     
An efficient way to manipulate the transport frequency is to tune the height of the potential barriers.   
Then, the tunneling frequency can be enhanced by using a shallow lattice as the corresponding energy gaps between the different subbands become smaller  
[compare Figs. \ref{fig:eigenspectra} ($a$), ($b$)], thereby making each tunneling process more favorable.   
Indeed, as shown in Fig. \ref{fig:fidelity_SF} ($e$) the single-particle tunneling modes for $V_0=4$ are located at larger frequencies $\omega''_{1,2}\approx 0.6, 1.2$ 
when compared to $V_0=10$ with the former being the most energetically favorable as it possesses the highest amplitude.    
On the other hand, the two particle transport modes, namely $SP\to S$ and $S\to T$, are negatively shifted with respect to their  
corresponding frequencies for $V_0=10$. 
The latter is a consequence of the decreasing subband energy gap occuring for shallower lattices, see also Fig. \ref{fig:eigenspectra}.   
Alternatively, a similar manipulation of the tunneling frequency can be achieved by tuning the postquenched state, by e.g. the value of $g_f$.  
We observe that for smaller (larger) $g_f$'s the corresponding tunneling branches are negatively (positively) shifted when compared to $g_f=2$ and $V_0=10$ [see Fig. \ref{fig:fidelity_SF} ($e$)].  
It is important to note here that the frequency peaks located at $\omega'''_2\approx 0.6, 1.1$ refer to tunneling between energetically different $SP$ states and the $S$ class.  
Summarizing, it has been shown that by tuning either the height of the potential barrier or the postquenched state via $g_f$ we can manipulate the location and the 
intensity of the tunneling frequency branches.  

Next we investigate the excitation dynamics, and in particular the time averaged probability of finding at least one boson within  
a single-particle state of a higher-band [see also Eq. (\ref{excitation})].  
Fig. \ref{fig:excitation_SF} ($a$) presents $\bar{P}_{exc}(\tau)$ for varying $\tau$. 
As it is evident and further confirmed by a direct numerical fit, the mean excitation probability obeys the bi-exponential law 
\begin{equation}
\label{scaling_1}
\bar{P}_{exc}(\tau)=Ae^{-\tau/\tau_1}+Be^{-\tau/\tau_2},  
\end{equation}
where $A$, $B$ are positive constants. 
Here, the two time scales introduced by the positive constants $\tau_1$, $\tau_2$ of the bi-exponential are necessary in order to describe accurately 
the mean excitation dynamics covering the sudden to adiabatic interaction quench regimes. 
The existence of these two time scales can be explained by the behavior of the number states that contribute to the MB state at each ramping interval.  
For convenience let us refer to the ramping time at the border between diabatic and moderate to nearly adiabatic ramping as $\tau_a$ [see also Fig. \ref{fig:excitation_SF} ($a$)]. 
Then, following the LIQ for $\tau<\tau_a$ where $\bar{P}_{exc}(\tau)$ exhibits a rapid decay, the avoided crossing located at $g=0$ in the eigenspectra is crossed 
diabatically and the MB state after the quench consists of a superposition of all lowest band number state classes. 
However for $\tau>\tau_a$, where $\bar{P}_{exc}(\tau)$ features a slow decay, the crossing at $g=0$ is traversed in a more adiabatic manner and the system after the quench 
is again in a superposition of the above mentioned states but now the state $\ket{1,1,1}$ possesses the main contibution.  
Alternatively, the above description can be explained by the fact that the $SP$ and $T$ categories for varying $\tau$ exhibit different scalings. 
Indeed, for $\tau<<\tau_a$ the population of the $SP$ increases while the $T$ slowly decreases for increasing $\tau$.  
However, for $\tau\ge\tau_a$ both categories decay in favor of the $S$ state. 
Having described the mechanism behind this bi-exponential decay of $\bar{P}_{exc}(\tau)$ let us demonstrate whether this law is robust also 
within the MF approximation or upon varying the initial state of the system. 
As it can be seen in Fig. \ref{fig:excitation_SF} ($a$) $\bar{P}_{exc}(\tau)$ exhibits the same bi-exponential law also within the MF approximation but the predicted 
amount of excitations is lesser when compared to the MB approach. 
Therefore, we can infer that the MF approximation predicts the qualitative decay of excitations but fails to capture the quantitative amount of excitations. 
Turning again to the MB approach, we observe that by initializing the system in a weakly interacting state $\bar{P}_{exc}(\tau)$ shows the same decay law 
but fewer excitations are produced than starting from $g_{i}=0$.  
This can be explained by the fact that the population of the $T$ category in the initial state is strongly suppressed for increasing $g_i$. 
Additionally, Fig. \ref{fig:excitation_SF} ($a$) shows $\bar{P}_{exc}(\tau)$ for a shallower lattice. 
$\bar{P}_{exc}(\tau)$ again exhibits a bi-exponential decay but most importantly it is observed that at $\tau\approx19$ the curves for $V_0=10$ and $V_0=4$ cross each other. 
As a consequence for $\tau<19$ [$\tau>19$] $\bar{P}_{exc}(\tau)$ is larger [smaller] for the shallower lattice. 
The above-mentioned crossing between the $\bar{P}_{exc}(\tau)$ for different heights of the potential barriers serves as a starting-point for investigating in the following the behavior 
of the mean excitation probability for varying $V_0$ and fixed $\tau$, see Fig. \ref{fig:excitation_SF} ($b$). 
A compatible double Gaussian fit $\bar{P}_{exc}(V_0)=A_1e^{-( \frac{V_0-C_1}{C_2})^2}+B_1e^{-(\frac{V_0-D_1}{D_2})^2}$  
(where $A_1$, $B_1$, $C_1$, $C_2$, $D_1$, $D_2$ refer to positive constants) is provided as a guide to the eye. 
As it can be seen, the scaling of $\bar{P}_{exc}(V_0)$ depends strongly on the considered ramping time $\tau$. 
Indeed, proximally to the diabatic limit, e.g. see $\tau$=1, $\bar{P}_{exc}(V_0)$ decreases for increasing $V_0$. 
Interestingly enough, for larger $\tau$, e.g. see $\tau$=25, $\bar{P}_{exc}(V_0)$ exhibits a completely different behavior. 
Initially it increases for increasing $V_0$ until it reaches a maximum value and then decreases for larger $V_0$. 
For instance, at $\tau=25$ $\bar{P}_{exc}(V_0)$ increases until $V_0\approx6$ where it exhibits a maximum and then decreases for $V_0>6$.  
We also remark here that the maximum of $\bar{P}_{exc}(V_0)$ is displaced for varying $\tau$, e.g. for $\tau=10$ is located at $V_0\approx5.5$ 
while for $\tau=50$ is at $V_0\approx8$. 
The above described alternating behavior of $\bar{P}_{exc}(V_0)$ for the different $\tau$ regions can be traced back to the previously 
observed crossing of $\bar{P}_{exc}(\tau)$ for different $V_0$, shown in Fig. \ref{fig:excitation_SF} ($a$) at $\tau=19$.   
Indeed, for $\tau<19$ ($\tau>19$) $\bar{P}_{exc}(\tau)$ decreases (increases) for increasing $V_0$, i.e. $\bar{P}_{exc}(\tau<19,V_0=4)>\bar{P}_{exc}(\tau<19,V_0=10)$. 
This is due to the fact that the fast (slow) decay process described by $\tau_1$ ($\tau_2$) being related to the $T \to SP$ ($SP \to S$) mode is stronger  
(weaker) for increasing $V_0$, see also Fig. \ref{fig:excitation_SF} ($a$). 
Comparing $\bar{P}_{exc}(V_0)$ for a fixed $V_0$ and varying $\tau$, we observe that $\bar{P}_{exc}(V_0)$ decreases in a uniform manner for increasing $\tau$,  
a result that is in accordance with our previous observations for the $\bar{P}_{exc}(\tau)$ bi-exponential decay, see Fig. \ref{fig:excitation_SF} ($a$). 
However, for deeper lattices ($V_0>12$) the decrease of $\bar{P}_{exc}(V_0)$ almost halts after a certain $\tau_0$ (mainly $\tau_0>10$). 

Finally, we study the impact of the postquench interaction strength on the higher-band excited fraction, see Fig. \ref{fig:excitation_SF} ($c$). 
$\bar{P}_{exc}(g_f)$ exhibits two distinct response regions with respect to $g_f$.  
These regions are classified by the location $g_C$ of the existing avoided crossings between the $SE$ and $T$ as well as $T$ and $HE$ categories in the corresponding MB 
eigenspectrum, see the grey scaled area which incorporate the solid elipses in Fig. \ref{fig:eigenspectra} ($a$).   
Indeed, when $g_f$ approaches $g_C$ $\bar{P}_{exc}(g_f)$ exhibits a non-linear growth as shown in the grey scaled interaction interval of Fig. \ref{fig:excitation_SF} ($c$), 
while for $g_f>g_C$ it increases in a linear manner. 
Overall, $\bar{P}_{exc}(g_f)$ grows for larger $g_f$'s as we import more energy to the system and thus excite more higher-band states.  
On the contrary, $\bar{P}_{exc}(g_f)$ reduces for larger $\tau$'s because for more adiabatic LIQs the effect of the 
existing avoided crossings is smeared out.  

In the next subsection we examine the negative LIQ dynamics, within the triple well system, from the MI to the SF correlated regimes. 
In particular, we shall elaborate in detail how such a negative LIQ alters the overall response of the system consisting 
of inter and intraband tunneling dynamics.

\subsection{Quench dynamics from MI to SF phase } \label{MOTTtoSF}

\begin{figure}[ht]
         \centering
            \includegraphics[width=0.5\textwidth]{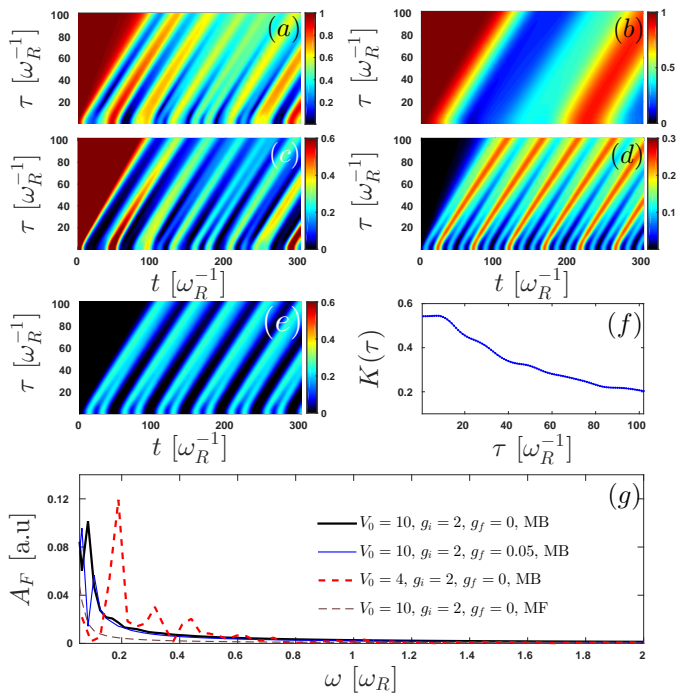}
                 \caption{Fidelity evolution for varying ramp time $\tau$ after a LIQ from $g_i=2$ to $g_f=0$. 
                Shown are the cases of ($a$) a shallow $V_0=4$ and ($b$) a deep $V_0=10$ triple well. 
                Probabilities of specific number state configurations during the evolution, namely ($c$) $P_1(t)=|<1,1,1|\Psi(t)>|^2$, ($d$) $P_2(t)=|<1,2,0|\Psi(t)>|^2=|<0,2,1|\Psi(t)>|^2$ 
                and ($e$) $P_3(t)=|<0,3,0|\Psi(t)>|^2$.    
                ($f$) Normalized variance of the fidelity $K(\tau)$ in the case of a shallow lattice for varying ramp time $\tau$.  
                ($g$) Spectrum of the fidelity following a LIQ with $\tau=8$ for different barrier heights, initial and final interactions 
                within the MB approach or the MF approximation (see legend).  
                The system consists of three initially strongly interacting bosons (see legend) confined in a triple well. }
                 \label{fig:fidelity_Mott}
 \end{figure}

The system is initialized within the strongly correlated regime, $g_i=2$, and therefore the ground state corresponds to the spatial contribution $\ket{1,1,1}$. 
To induce the dynamics we perform a negative LIQ to a weakly or non-interacting state and examine the dynamical response of the system. 
Fig. \ref{fig:fidelity_Mott} ($a$) presents $F(t;\tau)$ in a shallow triple well, $V_0=4$, for varying ramp time $\tau$ and $g_f=0$.  
As in the positive LIQ scenario $F(t)$ exhibits oscillations during the evolution possessing here, however, only a few small value frequencies. 
The fact that $F(t;\tau)<1$ indicates the melting of the MI phase \cite{Bernier}. 
The MB state after the quench consists of a superposition of the $S$, $SP$ and $T$ classes, see also Fig. \ref{fig:eigenspectra} ($a$).  
To justify the latter, we resort to the probability of specific number states 
that belong to the above classes.  
Indeed, Figs. \ref{fig:fidelity_Mott} ($c$), ($d$) and ($e$) present $P_1(t)=|\braket{1,1,1|\Psi(t)}|^2$, $P_2(t)=|\braket{1,2,0|\Psi(t)}|^2=|\braket{0,2,1|\Psi(t)}|^2$ and 
$P_3(t)=|\braket{0,3,0|\Psi(t)}|^2$ respectively. 
We observe that for diabatic LIQs the three modes are almost equally populated, while for large $\tau$'s the $SP$ mainly contributes to the dynamics and the other   
states possess a decaying amplitude in time. 
The reduced amplitude of $\ket{1,1,1}$ for larger $\tau$ indicates that the system is significantly perturbed 
for proximally adiabatic LIQs. 
To further elaborate on the system's dynamical response, Fig. \ref{fig:fidelity_Mott} ($f$) illustrates 
the corresponding normalized variance of the fidelity $K(\tau)$ for varying $\tau$. 
We observe an almost monotonic decrease of $K(\tau)$ for increasing $\tau$, especially for $\tau>10$, suggesting that the dynamical response 
is enhanced in the abrupt limit and reduces when the adiabatic limit is approached. 
The same overall response is also observed for a deeper triple well, see Fig. \ref{fig:fidelity_Mott} ($b$).   
However, the corresponding decaying response for 
increasing $\tau$ is more faint due to the increased energy gap between the MB eigenstates, see also Fig. \ref{fig:eigenspectra}.    

 \begin{figure}[ht]
         \centering
            \includegraphics[width=0.45\textwidth]{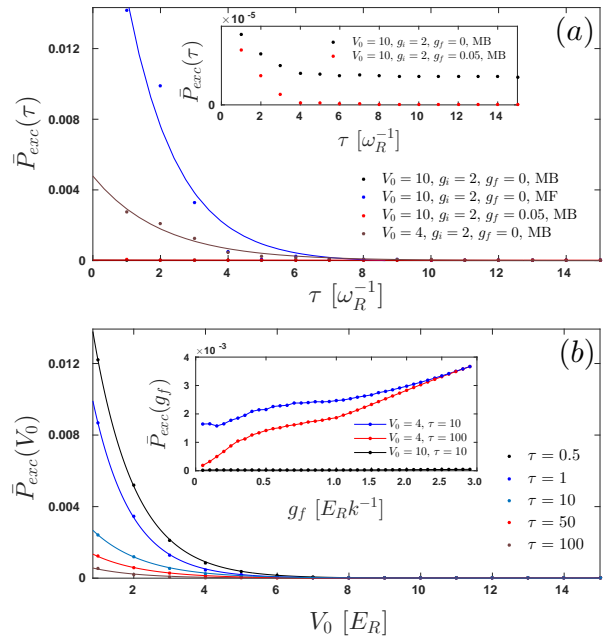}
                 \caption{Mean excitation probability of at least one boson to reside in an excited band for varying ($a$) ramp time $\tau$ of the LIQ 
                 (see a magnified version in the inset),    
                 ($b$) barrier height $V_0$ and in the inset of ($b$) postquench interaction strength $g_f$ with $g_i=3$. 
                 In all cases we consider a LIQ from an initially strongly to a final non correlated state unless stated otherwise (see legends).  
                 Different curves correspond to different parameter values (see legend), while the lines stem from a numerical fitting and 
                 provide a guide to the eye. }
                 \label{fig:excitation_Mott}
 \end{figure}

To investigate the tunneling dynamics we rely on the fidelity spectrum, see Fig. \ref{fig:fidelity_Mott} ($g$). 
It is observed that, in the case of a deep lattice ($V_0=10$), following a LIQ to $g=0$ only the $S \to SP$ tunneling mode is excited which is located at $\omega\approx0.1$. 
Employing the corresponding MF approximation this tunneling mode hardly survives with frequency $\omega'=0.001$ 
[hardly visible in Fig. \ref{fig:fidelity_Mott} ($g$)]. 
Turning again to the MB correlated approach, we investigate the influence on the tunneling dynamics of the postquench state and the dependence on the potential barrier. 
Examining first the situation of a weakly correlated postquench state, namely $g_f=0.05$, we observe that two lowest band tunneling modes exist. 
The first mode located at $\omega_1\approx0.02$ refers to the transport $S \to SP$ and the second at $\omega_2\approx0.11$ 
to tunneling $S \to T$.   
On the other hand, following a LIQ in a shallow lattice to a non-interacting final state mainly three tunneling modes are triggered.  
The first located at $\omega'_1=0.2$ refers to transport $S \to SP$, while the remaining two 
possess frequencies $\omega'_2=0.35$ and $\omega''_2=0.45$ which correspond to tunneling between the $S$ and different states belonging to the $T$ class.  
Note that higher frequency peaks, e.g. at $\omega_3\approx0.62$, refer to higher order lowest band transitions, 
for instance $SP \to T$, and possess reduced amplitudes.  

To complement our study, let us explore the mean amount of higher-band excitations induced by a negative LIQ. 
As in the previous section we examine the impact of the mean excitation dynamics as a function of the ramping time $\tau$ [see Fig. \ref{fig:excitation_Mott} ($a$)]  
or the height of the potential barrier $V_0$ [see Fig. \ref{fig:excitation_Mott} ($b$)].  
Here, we observe that the mean excitation dynamics for varying ramp time obeys an exponential decay
\begin{equation}
\label{scaling_3}
\bar{P}_{exc}(\tau)=A_3e^{-\tau/\tau_3},  
\end{equation}
where $A_3$, $\tau_3$ correspond to positive constants, and $\tau_3$ characterizes the rate of the decay.  
As shown the amount of produced excitations is negligible for all ramp times, see Fig. \ref{fig:excitation_Mott} ($a$).  
It reduces for a LIQ to a weakly correlated instead of the non-interacting state [see the inset of Fig. \ref{fig:excitation_Mott} ($a$)] and it is enhanced for shallower lattices. 
Finally, the corresponding MF approximation shows the same qualitative behavior but quantitatively fails to predict the correct amount of excitations. 
In addition, we find that the mean excitation dynamics follows again an exponential decay with respect to the barrier height, namely  
\begin{equation}
\label{scaling_4}
\bar{P}_{exc}(V_0)=A_4e^{-V_0/V_1}, 
\end{equation}
where $A_4$, $V_1$ refer to positive constants and $V_1$ is the inverse rate of the decay.  
As it can be deduced from Fig. \ref{fig:excitation_Mott} ($b$) again the produced amount of excitations is negligible and reduces even further when a  
larger ramp time is considered. 
Finally, let us examine the dependence of the excited to higher-band fraction on the postquench state for fixed ramping rate, 
shown in the inset of Fig. \ref{fig:excitation_Mott} ($b$). 
We observe that the population of excited states is overall negligible, and in particular it is greatly reduced when we enter the 
SF regime namely $g_f<0.5$ as well as in the case of deep optical lattices.  
Concluding, we can infer that the excitation dynamics following a LIQ from a MI-like to a weakly or even non-interacting final state is 
negligible. Therefore the lowest band approximation provides an adequate description of the system's dynamics.   

To demonstrate the robustness of our results for larger optical lattices, in the following section, we proceed to 
the investigation of unit filling systems which contain higher number of bosons confined in multiwell traps.    
In particular, we show that the dynamical response induced by a LIQ exhibits similar characteristics to the triple well case.

\section{LIQ Dynamics in extensive unit filling lattice systems} \label{fivewells}

Let us now investigate the dynamical response for larger unit filling setups.  
Here, we mainly focus on a five well optical lattice and consider a LIQ from a SF to a MI-like state and vice versa. 
Concerning the initial state of the system in the SF phase it consists of an admixture of the Wannier number states 
$\ket{0,1,3,1,0}$, $\ket{0,2,3,0,0}$, $\ket{0,2,2,1,0}$, $\ket{1,1,2,1,0}$, $\ket{1,1,1,1,1}$, while in the MI phase  
the dominant contribution stems from the $\ket{1,1,1,1,1}$ state. 
Note also that due to the underlying spatial symmetry of the system all the corresponding parity symmetric states contribute as well. 
 \begin{figure}[ht]
         \centering
            \includegraphics[width=0.45\textwidth]{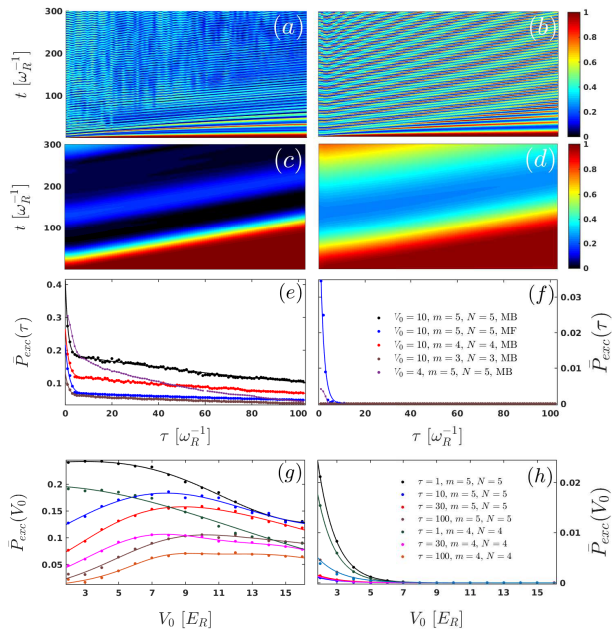}
                 \caption{Fidelity evolution for varying ramp time $\tau$ after a positive LIQ from $g_i=0$ to $g_f=2$ 
                 within ($a$) the MB approach and ($b$) the MF approximation. 
                 ($c$), ($d$) The same as ($a$), ($b$) but following a negative LIQ from $g_i=2$ to $g_f=0$. 
                 The system consists of five initially ($a$), ($b$) non and ($c$), ($d$) strongly interacting bosons confined in five wells.  
                 $\bar{P}_{exc}(\tau)$ for varying ramp time $\tau$ of the ($e$) positive and ($f$) negative LIQ.   
                 ($g$), ($h$) The same as ($e$), ($f$) but for varying barrier height $V_0$. 
                 Different curves correspond to different parameter values (see legends), while the lines stem from a numerical fitting and 
                 provide a guide to the eye. The legend shown in ($f$) [($h$)] is the same also for ($e$) [($g$)].}
                 \label{fig:fidelity_five}
 \end{figure}

To trigger the dynamics we employ either a positive or a negative LIQ protocol. 
Fig. \ref{fig:fidelity_five} ($a$) shows $F(t;\tau)$ following a positive LIQ for varying $\tau$. 
As in the triple well case, we observe that $F(t;\tau)$ exhibits an oscillatory behavior in time possessing also multiple 
frequencies which mainly correspond to lowest band transport and a few interband tunneling modes. 
The amplitude of the $F(t;\tau)$ oscillation becomes larger for increasing $\tau$ when referring to a fixed time instant $t$. 
This indicates that the system maximally departs from its initial state in the proximity of an adiabatic LIQ. 
Then by exploiting $\tau$ we can adjust the sweeping rate of the avoided crossings in the corresponding 
MB eigespectrum (not shown here for brevity) and thus control the population of the resulting excitations. 
Interestingly enough and in contrast to the triple well system here $\bar{F}(t>\tau;\tau)<0.3$, suggesting that larger unit filling setups 
can be driven out-of-equilibrium in a more efficient manner which is a manifestation of the Anderson orthogonality catastrophe \cite{Anderson1,Knap,Campbell}.   
As a consequence more modes can be triggered pointing also to the opportunity for enhanced interband tunneling (see below). 
To demonstrate the MB character of the dynamics we next present $F(t;\tau)$ employing the corresponding MF approximation, see Fig. \ref{fig:fidelity_five} ($b$). 
Despite the overall tendency for enhanced dynamical response for increasing $\tau$, $F(t;\tau)$ differs significantly from the MB approach.  
Indeed, a reduced number of modes is participating most of which refer to single-particle transport while $0.35<\bar{F}(t>\tau;\tau)<0.8$ in direct contrast to 
Fig. \ref{fig:fidelity_five} ($a$). 
The existing modes possess positive shifted frequencies, when compared to the MB approach. 

Next we examine the dynamical response induced by a negative LIQ, namely the MI melting, employing again $F(t;\tau)$, see Fig. \ref{fig:fidelity_five} ($c$).  
We observe that $F(t;\tau)$ performs oscillations of both large amplitude and period, while for larger $\tau$'s the system becomes more perturbed 
[hardly visible in Fig. \ref{fig:fidelity_five} ($c$)].    
After the LIQ the MB state consists of a superposition of several lowest band number states among which the states of double occupancy, e.g. $\ket{0,2,2,1,0}$, 
possess the main contribution and become dominant as we tend to an adiabatic ramping.   
The above result is in agreement with the negative LIQ dynamics of the triple well, however here $F(t;\tau)$ performs oscillations of both larger amplitude and period.  
Within the corresponding MF approximation, see Fig. \ref{fig:fidelity_five} ($d$), $F(t;\tau)$ shows the same overall qualitative decrease for larger $\tau$'s but 
quantitatively the dynamical response is altered significantly.  
For instance, $F(t;\tau)$ and its oscillation frequencies are larger when compared to the MB approach.   

Turning to the investigation of the induced interband tunneling we employ the mean excitation probability $\bar{P}_{exc}(t;\tau)$ for varying $\tau$.  
Following a positive [negative] LIQ $\bar{P}_{exc}(\tau)$ obeys a bi-exponential [an exponential] decay law, 
see Fig. \ref{fig:fidelity_five} ($e$) [Fig. \ref{fig:fidelity_five} ($f$)] similar to the triple well case.  
Note also that, as before, following a negative LIQ $\bar{P}_{exc}(\tau)$ is negligible.  
Additionally, $\bar{P}_{exc}(\tau)$ is enhanced for larger unit filling setups, while the MF approximation predicts a smaller (enhanced) amount of excitations for positive (negative) LIQs.  
Following a positive LIQ there is a crossing point (here at $\tau=8$) between $\bar{P}_{exc}(\tau)$'s, as in the triple well case, that refer to different $V_0$'s. 
This behavior of $\bar{P}_{exc}(\tau)$ seems to be quite robust in different setups and suggests the existence of two excitation time scales concerning 
the ramping rate and $V_0$.  
Namely, within a positive diabatic (adiabatic) LIQ the excited to higher-band fraction is larger (smaller) for shallower lattices. 
To complement our study on the excitation dynamics we investigate $\bar{P}_{exc}(V_0)$ for varying $V_0$ both for a positive LIQ, see Fig. \ref{fig:fidelity_five} ($g$), 
as well as a negative LIQ, see Fig. \ref{fig:fidelity_five} ($h$).  
We observe that within a positive LIQ scenario $\bar{P}_{exc}(V_0)$ strongly depends on $\tau$. 
In agreement to the triple well case, $\bar{P}_{exc}(V_0)$ decreases for small $\tau$ [e.g. see $\tau=1$ in Fig. \ref{fig:fidelity_five} ($g$)], while 
for larger rampings $\bar{P}_{exc}(V_0)$ shows a maximum at a specific region of $V_0$ [e.g. see $\tau=10$ in Fig. \ref{fig:fidelity_five} ($g$)].  
However, for a negative LIQ $\bar{P}_{exc}(V_0)$ follows an exponential decay with respect to $V_0$. 
Overall, in both positive and negative LIQs the excited to higher-band mean fraction increases for a more diabatic ramping and also for larger systems of unit filling. 
Finally, let us note that for the five well system $\bar{P}_{exc}(V_0)$ is not negligible suggesting that for more extensive unit filling lattices 
the occupation of higher-band states might be unavoidable even in the course of the negative LIQ dynamics.

\section{Conclusions and Outlook}\label{conclusions}

We have explored the nonequilibrium quantum dynamics following a linear interaction quench protocol in repulsively interacting few boson ensembles confined in 
finite optical lattices. 
The focus has been on unit filling such that the ground state of the system for increasing interaction strength exhibits the transition from a SF to a MI phase.   
To realize this transition and obtain the interaction dependence of the occupation of the number states, we first calculate the many-body eigenspectrum 
for varying interparticle repulsion.  
Here, the existence of multiple avoided crossings before and after the transition is elucidated. 
To induce the dynamics we perform a LIQ and cross dynamically, with a finite ramp rate, the aforementioned transition from both directions.  
Subsequently, we explore the dynamical response caused by the LIQ and in particular examine its dependence on several system parameters, 
such as the height of the potential barrier. 
It is important to note here that within our multiband treatment the dynamical response consists of both the lowest band tunneling and the 
excited to higher-bands fraction. 

Crossing the weak to strong interaction regimes yields the excitation of several lowest band tunneling pathways 
consisting of single and two particle transport. 
Furthermore, a rich interband tunneling dynamics is observed possessing mainly a single excitation to the first or second excited band. 
Analyzing in more detail the excited to higher-band fraction we examine its dependence on the quench ramp rate and barrier height. 
We find that it exhibits a bi-exponential decay for decreasing quench rate. 
This decay law introduces two different time scales in the excitation dynamics, which are directly related to the diabatic 
or adiabatic crossing of the transition respectively and can be further explained by the behavior of the participating number states.  
Furthermore, the excited fraction follows a more complex scaling for varying height of the potential barrier.   
For diabatic quenches it reduces, while for larger ramp times it exhibits a non-linear behavior. 
The latter can be explained by exploiting the dependence on the ramp time of the excited fraction for shallow and deep lattices. 
Finally, the higher-band dynamics strongly depends on the postquench state, namely when we tend to the region of an existing avoided crossing 
it is described by a non-linear growth, while for larger quench amplitudes it increases in an almost linear manner. 

In constrast to the above, the dynamical response following a LIQ from strong to weak interactions is reduced and mainly 
comprises of the lowest band tunneling dynamics. 
Indeed, in this case following the quench we can excite only a few tunneling modes, while the excited to higher-band fraction is negligible and obeys 
an exponential decay for varying ramp time. 
Here, the lowest band approximation seems to describe the induced dynamics accurately.  
Finally, we made an attempt to generalize our results by considering larger systems, e.g. a five well setup, and showing the robustness of the above 
mentioned scalings as well as the enhancement of the excited to higher-band fraction. 

Finally, let us comment on possible future extensions of our work. 
An interesting alternative of the present work would be to investigate the dynamical response induced by a LIQ in repulsively interacting dipolar bosons \cite{Chatterjee3}  
upon crossing the corresponding SF to supersolid transition point.  
Certainly, the study of bosonic or fermionic spinor ensembles confined in an optical lattice is an intriguing perspective.   
Here, the inclusion of the spin degree of freedom enriches the phase diagram \cite{Lewenstein,Tsuchiya,Rizzi} and 
as a consequence it might alter significantly the dynamical response.

\appendix

\section{The Computational Approach: MCTDHB}\label{appendix}

To solve the many-body (MB) Schr\"{o}dinger equation $\left( {i\hbar {\partial _t} - H} \right)\ket{\Psi (t)} = 0$ 
of the interacting bosons as an initial value problem $\ket{{\Psi (0)}} = \left| {{\Psi _0}}
\right\rangle$, we rely on the Multi-Configuration Time-Dependent Hartree method for Bosons (MCTDHB) \cite{Alon,Alon1,Streltsov}. 
The latter has already been applied for a wide set of nonequilibrium bosonic settings, e.g. see \cite{Streltsov,Streltsov1,Alon2,Alon3,Jannis,Mistakidis,Mistakidis1,Mistakidis2,Koutentakis,Katsimiga}. 
This method allows for a variationally optimal truncation of the Hilbert space as we employ a time-dependent   
moving basis where the system can be instantaneously optimally represented by time-dependent permanents. 
The MB wavefunction is expanded in terms of the bosonic number states $\left| {{n_1},{n_2},...,{n_M};t}\right\rangle$, 
that built upon time-dependent single-particle functions (SPFs) $\left| \phi_{i}(t) \right\rangle$, $i=1,2,...,M$, and time-dependent weights $C_{\vec{n}}(t)$ 
\begin{equation}
\label{eq:10}\left| {\Psi (t)} \right\rangle  = \sum\limits_{\vec n
} {{C_{\vec n }}(t)\left| {{n_1},{n_2},...,{n_M};t} \right\rangle }.
\end{equation}
Here $M$ is the number of SPFs and the summation $\vec n$ is over all the possible combinations $n_{i}$ such that the total number
of bosons $N$ is conserved. 
Note that in the limit in which $M$ approaches the number of grid points the above expansion is equivalent to a full configuration interaction approach.   
Furthermore, in the case of $M=1$ the MB wavefunction is given by a single permanent $\ket{n_{1}=N;t}$ and the method reduces to the 
time-dependent Gross Pitaevskii mean-field approximation. 

To determine the time-dependent wave function $\left|\Psi(t) \right\rangle$ we calculate the equations of motion for the coefficients 
${{C_{\vec n }}(t)}$ and the SPFs $\left| \phi_{i}(t) \right\rangle$.
Following the Dirac-Frenkel \cite{Frenkel,Dirac} variational principle, ${\bra{\delta\Psi}}{i{\partial _t} - \hat{ H}\ket{\Psi }}=0$, 
we obtain the well-known MCTDHB equations of motion \cite{Alon,Streltsov,Alon1}.  
These equations consist of a set of $M$ non-linear integrodifferential equations of motion for the SPFs being coupled to the 
$\frac{(N+M-1)!}{N!(M-1)!}$ linear equations of motion for the coefficients ${{C_{\vec n }}(t)}$. 
Finally, let us remark that in terms of our implementation we 
use an extended version of MCTDHB being referred to in the literature as the Multi-Layer Multi-Configuration 
Time-Dependent Hartree method for bosonic and fermionic Mixtures (ML-MCTDHX) \cite{Cao_ML,Kronke,Cao_X}. 
This computational package is particularly suitable for treating systems consisting of different bosonic, fermionic species, 
while for the case of a single bosonic species it reduces to MCTDHB. 

For the numerical implementation, the SPFs are expanded within a so-called primitive basis $\lbrace \left| k \right\rangle \rbrace$ of dimension $M_{p}$. 
As a primitive basis for the SPFs we have used a sine discrete variable
representation, which intrinsically introduces hard-wall boundaries 
at both ends of the potential.  
To obtain the $n$-th MB eigenstate we rely on the so-called improved relaxation scheme, being summarized as follows.   
First, we initialize the system with an ansatz set of SPFs $\lbrace |\phi_i^{(0)} \rangle \rbrace$, diagonalize the Hamiltonian within a basis spanned by the SPFs 
and set the $n$-th obtained eigenvector as the $C_{\vec{n}}^{(0)}$-vector.  
Then, we propagate the SPFs in imaginary time within a finite time interval $d \tau$, update the SPFs to $\lbrace |\phi_i^{(1)} \rangle \rbrace$ and 
repeat the above steps until the energy of the state converges within the prescribed accuracy. 
In turn, we perform a time-dependent quench on the strength of the interparticle repulsion and study the
evolution of $\ket{\Psi (t)}$ in the $m$-well potential by utilizing the appropriate 
Hamiltonian within the MCTDHB equations of motion. 

To track the numerical error and guarantee the accurate performance of the numerical integration for the MCTDHB 
equations of motion we impose the following overlap criteria $|\langle \Psi |\Psi \rangle -1| < 10^{-9}$ and
$|\langle \varphi_i |\varphi_j \rangle -\delta_{ij}| < 10^{-10}$ for the total wavefunction and the SPFs respectively. 
The dimension of the used primitive basis consists of 300 spatial grid points in the case of a triple well and 
500 spatial grid points for the five well potential. 
Furthermore, to ensure the convergence of our simulations we have used up to 9 (10) optimized single
particle functions for the triple (five) well, thereby observing a systematic convergence of our results. 
An auxilliary indicator for convergence is provided by the population of the lowest occupied natural orbital kept always below $0.1\%$.

\section*{Acknowledgments}
The authors gratefully acknowledge funding by the Deutsche Forschungsgemeinschaft (DFG) in the framework of the
SFB 925 ''Light induced dynamics and control of correlated quantum
systems'' and by the excellence cluster ''The Hamburg Centre for Ultrafast Imaging-Structure, Dynamics and Control of Matter at the Atomic Scale''.

{}

\end{document}